\documentstyle[12pt]{article}

\def\numberbysection{\@addtoreset{equation}{section}
\def\theequation{\thesection.\arabic{equation}}}
\def\beq{\begin{equation}}
\def\eeq{\end{equation}}
\def\ra{\rightarrow}
\def\om{\omega}

\def\nn{\nonumber}
\def\eps{\epsilon}
\def\ga{\gamma}
\def\si{\sigma}
\def\de{\delta}
\def\pa{\partial}
\def\th{\theta}

\def\la{\lambda}

\begin{document}
\begin{center}
\hfill SNUTP-93/42\\
\vskip 0.3in
{\bf Coherent State Quantization of\\
SU(N) Non-Abelian Chern-Simons Particles}
\vskip 0.3in
Taejin Lee\\
{\em Department of Physics, Kangwon National University}\\
{\em Chuncheon 200-701, KOREA }\\
{E-mail:taejin@cc.kangwon.ac.kr}\\
\vskip 0.2in
Phillial Oh\\
{\em Department of Physics, Sung Kyun Kwan University}\\
{\em Suwon 440-746, KOREA}\\
{E-mail:ploh@yurim.skku.ac.kr}\\
\end{center}
\vskip 0.3in

\begin{abstract}
We present a new method of formulating the classical theory of $SU(N+1)$
non-Abelian Chern-Simons (NACS) particles for arbitrary $N\geq 1$
using the symplectic reduction of
$CP(N)$ manifold from $S^{2N+1}$. Quantizing  the theory
using BRST formulation and coherent state path integral method
, we obtain a quantum mechanical
model for $SU(N+1)$ NACS particles.

\end{abstract}
\newpage

The anyons,  particles carrying  anomalous
spin and exihibiting fractional statistics in two spatial
dimensions are an interesting theoretical curiosity
and  may be realized in condensed matter physics \cite{wil}.
The  anyons can be described as  particles carrying both  charge and
magnetic flux and an action for them can be
 constructed by coupling their charges minimally with
the Abelian Chern-Simons gauge field.
One can generalize the notion of
anyon by introducing a non-Abelian gauge group, {\it i.e.},
 particles
which carry non-Abelian charges and interact with each other through the
non-Abelian Chern-Simons term \cite {des,witt}.
We will call
them non-Abelian Chern-Simons (NACS) particles . One can
also construct
the corresponding classical action for the NACS particles
by introducing isospin degrees of freedom and coupling isospin charge
with the
non-Abelian Chern-Simons gauge fields \cite{bal}.
Recently, a derivation of quantum mechanical model for $SU(2)$ NACS
particles
from a classical action has been achieved in ref.\cite{lo}.
The resulting quantum mechanics showed that the NACS particles also carry
anomalous spins and satisfy yet more generalized braid statistics.
In this letter, we present a new method of formulating classical theory
of $SU(N+1)$ NACS particles with arbitrary $N\geq 1$ and
 quantize the theory using coherent state path integral \cite{klauder},
and thereby  obtain a quantum mechanical model for NACS particles.

A conventional way of formulating a classical action
for $SU(N+1)$ NACS particles is to construct
a first order  Lagrangean for internal degrees of freedom defined
on the $SU(N+1)$ group manifold and to introduce
isospin charges which transform under the adjoint representation of
the group \cite{bal2}.
Then, one minimally couples the isospin charges
 with non-Abelian Chern-Simons gauge field \cite{bal}.
It turns out that both first and second class constraints
\cite{dirac} arise from this classical Lagrangean for NACS particles
\cite{bal2,oh} and the quantization of the classical action for NACS
particles  is rather involved.

One way of remedying such complications would be to start with a
Lagrangean in which second class constraints do not appear
from the beginning if we suitably define the Poisson bracket \cite{jack}.
For that purpose, it would be better to consider $CP(N)$
manifold \cite{grif} instead of $SU(N+1)$ manifold
to describe internal degrees of freedom.
$CP(N)$ manifold is one of the coadjoint orbits
\cite{kril} of $SU(N+1)$ group and it can be
reduced from $S^{2N+1}$ via the
symplectic reduction \cite{gui} ,{\it i.e.},
$CP(N) \simeq S^{2N+1}/U(1)$.
In terms of a complex column vector $Z=(Z_0,Z_1,\cdots,Z_N)^T$ and
its complex conjugate
$\bar Z=({\bar Z}_0,{\bar Z}_1,\cdots ,{\bar Z}_N)$, $S^{2N+1}$ is defined
by the constraint
\beq
\phi=\bar  Z Z-1= 0.
\label{sphere}
\eeq
Now, let us consider first order Lagrangian
\beq
L_Z=i J(\bar Z\dot Z -\dot{\bar Z}Z)-h(Z,\bar Z)-\la\phi
\label{lag1}
\eeq
where $J$ is a constant and $\la$ is a Lagrange multiplier.
{}From this first order Lagrangian,  the Poisson bracket can be read
as follows:
\beq
\{F,G\}=-\frac{i}{2J}\sum_I\left(\frac{\pa F}{\pa {\bar Z}_I}
\frac{\pa G}{\pa Z_I}
-\frac{\pa F}{\pa Z_I}\frac{\pa G}{\pa {\bar Z}_I}\right)
\eeq
and the fundamental commutators are given by
\beq
\{{\bar Z}_I,Z_K\}= -\frac{i}{2J}\de_{IK},\ \ \ \ \ \ \{{\bar Z}_I,
{\bar Z}_K\}=
\{Z_I,Z_K\}=0.\label{poi1}
\eeq

The isospin charges are defined as
\beq
Q^a=2J\bar ZT^aZ\ \ \ \ \ \ \ (a=1,2,\cdots, N^2+2N)\label{def}
\eeq
where $T^a$ are the $(N+1)\times (N+1)$ matrices of $SU(N+1)$  satisfying
$[T^a,T^b]=if^{abc}T^c$ and Tr$(T^aT^b)=\frac{1}{2}\de_{ab}$.
Using Eq.(\ref{poi1}), we can check that
$Q^a$ satisfy $SU(N+1)$  algebra;
\beq
\{Q^a,Q^b\}=-f^{abc}Q^c.\label{op}
\eeq
Note that the theory described by the Lagrangean
Eq.(\ref{lag1}) is invariant
under $U(1)$ gauge
transformations generated by $\de f=\eps\{\phi,f\}$
provided  $\{\phi,h\}= 0$. In our case, $h(Z, \bar Z)$ turns out to be
a function of $Q^a$'s only and satisfies the condition due to
$\{\phi,Q^a\}=0$.

The next step is to introduce standard coordinates on $CP(N)$ by
$\xi_i=Z_i/Z_0(Z_0\neq 0, i=1,2,\cdots,N)$ and to make coordinate
transformation from $Z_I$ to $(Z_0,\xi_i)$. To reduce the internal group
manifold to $CP(N)$, we must choose a gauge condition and we find the
following gauge fixing is convenient  \beq
\chi=\frac{1}{2}({\bar Z}_0 -Z_0) = 0\label{gauge1}.
\eeq
Then, the solution to the constraint Eq.(\ref{sphere}) is given by
\beq
{\bar Z}_0=Z_0=\frac{1}{\sqrt{1+\mid \xi\mid^2}},\ \ \ \ \mid\xi\mid^2
=\sum_i\mid\xi_i\mid^2.
\label{sol1}
\eeq
By sustituting $Z_I=(Z_0,Z_0\xi_i)$ and Eq.(\ref{sol1}) into the
Eq.(\ref{lag1}), we obtain the following Lagrangean which describes the
particle motion in the internal space $CP(N)$, \beq
L=i J\frac{{\bar \xi}\dot \xi-\dot{\bar \xi}\xi}
{1+\mid\xi\mid^2}- h'(\xi,{\bar \xi})
\eeq
where $h'(\xi, \bar \xi)=h(Z,\bar Z)\mid_{{\bar
Z}_0=Z_0=1/\sqrt{1+\mid\xi\mid^2} ,Z_i=\xi_i Z_0}$.

Note that the first term in the above Lagrangean is of
the form $\int \th$
where the one form $\th$ is given by
\beq
\th=i J\frac{{\bar \xi}d\xi-d{\bar \xi}\xi}
{1+\mid\xi\mid^2}.
\eeq
It yields the well-known symplectic two form $\om$ on $CP(N)$,
\beq
\om=d\th=2i\left[J\frac{d{\xi}\wedge d{\bar\xi}}{1+
\mid\xi\mid^2}-\frac{({\bar \xi}d\xi)\wedge(\xi d{\bar \xi})}{(1
+\mid\xi\mid^2)^2}\right],
\eeq
and the Fubini-Study metric $ds^2=\sum_{i,k}g_{ik}d\xi_id{\bar \xi}_k$,
\beq
g_{ik}=\frac{2J}{(1+\mid\xi\mid^2)^2}[(1+\mid\xi\mid^2)\de_{ik}-
{\bar \xi}_i\xi_k],
\eeq
whose inverse $g^{ki}$ satisfying $g_{ik}g^{kj}=\de_i^j$ is given by
\beq
g^{ki}=\frac{1}{2J}(1+\mid\xi\mid^2)
(\de_{ki}+{\bar \xi}_k\xi_i).\label{metric}
\eeq

Using the Fubini-Study metric, we can define Poisson bracket on $CP(N)$
and the corresponding fundamental commutators as follows
\[\{F,G\}=-i\sum_{i,k}g^{ki}\left(\frac{\pa
F}{\pa {\bar \xi}_k}\frac{\pa G}{\pa \xi_i}-
\frac{\pa F}{\pa\xi_i}\frac{\pa G}{\pa {\bar \xi}_k}\right),\]
\beq
\{{\bar \xi}_i,\xi_j\}=-\frac{i}{2J}(1+\mid\xi\mid^2)
(\de_{ij}+{\bar \xi}_i\xi_j),\label{poi3}
\eeq
\[\{\xi_i,\xi_j\}=\{{\bar \xi}_i,{\bar \xi}_j\}=0.\]
The isospin charge $Q^a$ of Eq.(\ref{def}) can be re-expressed in terms of
coordinates $\xi_i$
\beq
Q^a=2J\sum_{I=0}^{N}{\bar Z}_IT^a_{IK}Z_K\mid_{Z_0=\frac{1}
{\sqrt{1+\mid\xi\mid^2}}, Z_i=Z_0\xi_i}.\label{def2}
\eeq
For  $SU(2)$ group with Pauli matrices $T^a=\si^a/2$, we have
\beq
Q^1=\frac{J(\xi+{\bar \xi})}{1+\mid\xi\mid^2},\ \ \ \ Q^2=i\frac{J
({\bar \xi}-\xi)}{1+\mid\xi\mid^2},\ \ \ \
Q^3=\frac{J(1-\mid\xi\mid^2)}{1+\mid\xi\mid^2}
\eeq
We can easily check that the isospin charges satisfy the same
$SU(N+1)$ algebra Eq.(\ref{op}) after the reduction.

Now we can write down the Lagrangean for a system of
$N_p$ isospin particles
which are interacting with $SU(N+1)$ Chern-Simons gauge field.
Denoting the spatial
and internal coordinates of the particles by
${\bf q}_\alpha$ and $\xi^\alpha_i$,
$\alpha=1,2,\cdots,N_p,\ \ i=1,2,\cdots,N$
, we have
\[L=L_{ptl}+L_{int}+L_{C-S},\]
\beq
 L_{ptl} = \sum_\alpha\left(-{1 \over 2} m_\alpha \dot{\bf
q}_\alpha^2 +
i J_{\alpha}\frac{{\bar \xi}^\alpha\dot \xi^\alpha-
\dot{\bar \xi^\alpha}\xi^\alpha}
{1+\mid\xi^\alpha\mid^2}\right),
\eeq
\[L_{int}=\int d^2{\bf
x}\sum_\alpha \left({\bf A}^a_\alpha(t,{\bf x})\cdot \dot{\bf q}_\alpha +
A^a_0(t,{\bf x})\right) Q^a_\alpha \delta ({\bf x}-{\bf q}_\alpha)
\label{lag},\]
\[L_{C-S}=\kappa\int
d^2 {\bf x} \,\epsilon^{\mu\nu\lambda} {\rm tr}\left(A_\mu \partial_\nu
A_\lambda +{2\over 3} A_\mu A_\nu A_\lambda\right), \]
where $\kappa=\frac{k}{4\pi}$  with $k$=integer,
$A_\mu=A_\mu^a T^a$ and the
space-time signature is $(+,-,-)$. The equations of motion obtained from
the above Lagrangean contain the Wong's equation \cite{wong}.

It is convenient to employ the complex coordinates, $z = x+ iy$, $\bar z
= x- iy$, $z_\alpha = q^1_\alpha + i q^2_\alpha$, $\bar z_\alpha =
q^1_\alpha
- i q^2_\alpha$, $A^a_z = {1\over 2} (A^a_1 - iA^a_2)$, $A^a_{\bar z} =
{1\over 2} (A^a_1 + iA^a_2)$ when we proceed to constraint analysis and
quantization. Introducing momenta variables, $p^{\bar z}_\alpha$,
$p^z_\alpha$ and $\pi^a$ which are canonical conjugates
to $z_\alpha$, $\bar
z_\alpha$ and $A^a_0$ respectively,  we obtain the following first order
Lagrangean
\[L = \sum_\alpha\left(p^{\bar z}_\alpha \dot z_{\alpha}+
p^{z}_\alpha \dot{{\bar z}}_{\alpha}+ i
J_{\alpha}\frac{{\bar \xi}^\alpha \dot \xi^\alpha-\dot{\bar
\xi^\alpha}\xi^\alpha}{1+\mid\xi^\alpha\mid^2}\right)\]
\beq
+\int d^2 z \left(\pi^a \dot A^a_0+
{\kappa\over 2}\left(\dot A^a_z A^a_{\bar z} -\dot A^a_{\bar z} A^a_{
z}\right) +  A^a_0 \Phi^a \right) - H,\label{flag}
\eeq
\[H = \sum_\alpha {2\over
m_\alpha}\left(p^{\bar z}_\alpha-A^{a}_z(z_\alpha, \bar z_\alpha)
Q^a_\alpha\right) \left(p^{z}_\alpha-A^{a}_{\bar z}
(z_\alpha, \bar z_\alpha)
Q^a_\alpha\right).\]
In passing, we note that we do not introduce conjugate momenta for
the variable which already appear as phase space variables.
Here $\Phi^a$'s are the Gauss constraints given by
\beq
\Phi^a(z)=\kappa F^a_{z\bar z}+\sum_\alpha Q^a_\alpha\de(z-z_\alpha)=0.
\label{cons}
\eeq

The above first order Lagrangean Eq.(\ref{flag}) leads us to define the
Poisson bracket as in ref.\cite{lo} along with Eq.(\ref{poi3}).
As expected, we find that
$\Phi^a$'s satisfy the $SU(N+1)$ algebra  and no further
constraints arise
\beq
\{\Phi^a(z),\Phi^b(z^\prime)\}=-f^{abc}\Phi^c \delta(z-z^\prime),\quad
\{H, \Phi^a(z)\}=0.
\eeq

If the Gauss' constraint is solved explicitly, dynamics of the NACS
particles can be described solely by the qunatum mechanical Hamiltonian
$H$ in Eq.(\ref{flag}). Thus it is desirable to choose a gauge condition
where the Gauss' constraint can be solved explicitly. Taking advantage of
the nature of 2+1 dimensions, one may realize that the axial gauge works.
Choosing an axial gauge condition, say, $A^a_1=0$,
we get a solution for the Gauss' constraint
\beq
A^a_1({\bf x})=0,\quad A^a_2({\bf x})=-\frac{1}{\kappa}\sum_\alpha
Q^a_\alpha\theta(x-x_\alpha)\delta(y-y_\alpha)+f(y)
\eeq
where $f(y)$ is an arbitrary function of $y$. Unfortunately, it is rather
awkward to describe the dynamics of the NACS particles
with this axial gauge
solution because of the strings attached to the particles.

A better gauge condition can be found if we adopt
the coherent state quantization \cite{ems,bos} for the gauge fields.
Adopting the coherent state quantization
effectively  enlarges the gauge orbit space
in that $A^a_z$ and $A^a_{\bar
z}$ are treated as independent variables. This enable us to choose
$A^a_{\bar z} = 0$ as  a gauge condition \cite{lo} in the
framework of the coherenet state quantization.
Since the gauge fields have only holomorphic parts in this gauge, it may be
called holomorphic gauge.
In this gauge the solution for the Gauss' constraint is
obtained \cite{lo,gua} as
\begin{equation}
A^a_{\bar z} (z, \bar z)= 0,\quad
A^a_z (z, \bar z) = {i\over 2\pi \kappa}\sum_\alpha  Q^a_\alpha
{1\over z -z_\alpha}.\label{sol}
\end{equation}

The BRST invariant physical transition amplitude is represented by a path
integral
\[
Z_\Psi=\int D p^z D q^{\bar z} D p^{\bar z} D q^z D \mu(\xi, \bar\xi)
D A_z D A_{\bar z} D \pi D A_0 D b D c D \bar b D \bar c \]
\begin{equation}
\exp\left\{
-\kappa i \int d^2z\left(A^f_{\bar z} A^f_z+A^i_{\bar z}
A^i_z\right)\right\}
\exp\left\{ i\int^{t_f}_{t_i} dt\left(K+\{\Psi, \Omega\}+
\int d^2 z(\dot c^a \bar b^a+ \dot b^a \bar c^a)\right)\right\}
\end{equation}
where $K= \sum_\alpha\left(p^{\bar z}_\alpha \dot z_{\alpha}+
p^{z}_\alpha \dot{{\bar z}}_{\alpha}+ i
J_{\alpha}\frac{{\bar \xi}^\alpha \dot \xi^\alpha-\dot{\bar
\xi^\alpha}\xi^\alpha}{1+\mid\xi^\alpha\mid^2}\right)
+\int d^2 z \left(\pi^a \dot A^a_0+
{\kappa\over 2}\left(\dot A^a_z A^a_{\bar z} -
\dot A^a_{\bar z} A^a_{z}\right) \right) - H$.
Also $\Omega$ is the nilpotent BRST charge defined by
\begin{equation}
\Omega = \int d^2 z(c^a\Phi^a-ib^a \pi^a -
\frac{1}{2} f^{abc}c^a c^b \bar b^c)
\end{equation}
and $\Psi$ is the Grassmann odd gauge function whose explicit
expression depends on the gauge condition chosen.
It should be understood that it is the BRST invariance which
gurantees the
the equivalence of the path integral in the holomorphic gauge to those
in more conventional gauges such as Coulomb and
axial gauges: $Z_\Psi$ is independent of $\Psi$ \cite{hen}.

In order to impose the holomorphic gauge condition, we take
\begin{equation}
\Psi= \int d^2 z \left(\frac{i}{\beta} \bar c^a A^a_{\bar z}+ \bar b^a
A^a_0\right)
\end{equation}
where $\beta$ is a parameter to be taken zero at the end.
Integrating out the BRST ghosts and the gauge fields as in ref.\cite{lo},
we end up with a quantum
mechanical description of the NACS particles:
\[Z=\int Dp^{\bar z} Dq^z Dp^z Dq^{\bar z} D\mu({\bar
\xi},\xi)\]
\beq
\times\exp\left\{i\int_{t_i}^{t_f}dt
\left(\sum_\alpha\left(p^{\bar
z}_\alpha \dot z_{\alpha}+ p^{z}_\alpha \dot{{\bar z}}_{\alpha}+
i J_{\alpha}\frac{{\bar \xi}^\alpha\dot \xi^\alpha-
\dot{\bar \xi^\alpha}\xi^\alpha}
{1+\mid\xi^\alpha\mid^2}\right)-H\right)\right\}\label{path}
\eeq
\[H=\sum_\alpha \frac{2}{m_\alpha}p_\alpha^z\left(p^{\bar z}_\alpha-
{i\over 2\pi \kappa}\sum_\beta  Q^a_\beta
{1\over z_\alpha -z_\beta}\right)\]
where $D\mu({\bar \xi},\xi)$ is the Liouville
measure on $CP(N)$ given by
\beq
D\mu({\bar \xi},\xi)=\prod_\alpha
D\mu({\bar \xi^\alpha},\xi^\alpha)=\prod_\alpha\frac{c\mid
d\xi^\alpha\mid}{(1+\mid\xi^\alpha\mid^2)^{N+1}}
\eeq
where $c$ is a normalization constant \cite{sch}.

Now, we only need to show that the above path integral is equal to the
following transition amplitude governed
by the Hamiltonian operator $\hat H$
\beq
Z = <\xi_F, \zeta_F\vert \exp\{ -i \hat{H}(t_F-t_I)\} \vert\zeta_I, \xi_I>
\label{prop}
\eeq
where $\zeta$ and $\xi$ denote collectively
$(z_\alpha, \bar z_\alpha)$ and
$(\xi_\alpha, \bar \xi_\alpha)$, and
$\mid \zeta,\xi>=  \prod_\alpha\mid z_\alpha>\otimes\mid \bar z_
\alpha>\otimes\mid \xi^\alpha>\otimes\mid \bar \xi^\alpha>$.
The Hamiltonian operator $\hat{H}$ is
the operator version of the classical
Hamiltonian Eq.(\ref{path}).
We achieve this by using coherent states path integral on $CP(N)$
\cite{klauder}.

We first construct coherent states on $CP(N)$. Let us consider
$\mid 0>$, the highest weight state annihilated by all positive roots
of $SU(N+1)$ algebra in Cartan basis.
Then for $CP(N)$ with given $P_\alpha
\equiv 2J_\alpha$ $(P_\alpha\in {\bf Z}^+)$ we have an irreducible
representation  $(P_\alpha,0,\cdots,0)$ of $SU(N+1)$ group according
to Borel-Weil-Bott theorem \cite{kril} and there are precisely
$N$ negative roots
$E_\ga,\ga=1,2,\cdots,N$ such that $E_\ga\mid 0>\neq\mid 0>$ for
the irreducible representations $(P_\alpha,0,\cdots,0)$. Let us
label $\{E_\ga\}=\{E_i\}$.
Representing a point on $CP(N)$ by $\xi^\alpha\equiv(\xi^\alpha_1,
\xi^\alpha_2,\cdots,
\xi^\alpha_N)$, we construct a
coherent state on $CP(N)$ by \cite{klauder}
\beq
\mid P_\alpha,\xi^\alpha>=\frac{1}{(1+\mid\xi^\alpha\mid^2)^{J_\alpha}}
\exp(\sum_i\xi^\alpha_iE_i)\mid 0>.
\eeq

Having defined the coherent states on $CP(N)$, we find that they have the
following two important properties which are necessary ingredients to take
the next step forward. One is the resolution
of unity, \beq \int D\mu({\bar \xi}^\alpha,\xi^\alpha) \mid
P_\alpha,\xi^\alpha> <P_\alpha,\xi^\alpha\mid=I,\label{id}
\eeq
and the other is reproducing kernel,
\beq
<P_\alpha,\xi^{\alpha\prime}\mid P_\beta,\xi^\beta>=
\frac{(1+\bar\xi^{\alpha\prime}\xi^\alpha)^{2J_\alpha}}
{(1+\mid\xi^{\alpha\prime}\mid^2)^{J_\alpha}
(1+\mid\xi^\alpha\mid^2)^{J_\alpha}}
\de_{\alpha\beta}.\label{kerr}
\eeq

Now we are in a position to show the equivalence between Eqs.(\ref{path})
and (\ref{prop}). Since the part of the spatial degrees of freedom can be
treated by canonical way, we only need to prove the equivalence in the
sector of the internal degrees of freedom. Consider \beq
Z=<P,\xi_F\mid \exp\{-i{\hat H}(t_F-t_I)\}\mid P,\xi_I>
\eeq
where $\mid P,\xi>=\prod_\alpha \mid P_\alpha,\xi^\alpha>$.
Let us divide the time interval by $M+1$ steps and $\eps=\frac{t_f-t_i}
{M+1}$. Also $\mid P,\xi_I>\equiv\mid P,\xi(0)>$ and $\mid P,\xi_F>\equiv
\mid P,\xi(M+1)>$. Inserting the identity  Eq.(\ref{id})
repeatedly, we have
\[Z=\lim_{M\ra\infty}
\int\cdots\int\prod_\alpha\prod_{n=1}^M D\mu({\bar
\xi}^\alpha(n),\xi^\alpha(n)) \prod_{n=1}^{M+1}
<P,\xi(n)\mid P,\xi(n-1)>\]
\beq
\times\left(1-i\eps\frac{<P,\xi(n)\mid {\hat H}\mid
P,\xi(n-1)>}{<P,\xi(n)\mid P,\xi(n-1)>}\right).
\eeq
Using the Eq.(\ref{kerr}) with $\xi^\alpha(n-1)=
\xi^\alpha(n)-d\xi^\alpha(n)$, we find
\begin{eqnarray}
\prod_{n=1}^{M+1} &<&P,\xi(n)\mid P,\xi(n-1)>\nn\\
&=&\prod_\alpha\prod_{n=1}^{M+1}
\left(1-J\frac{({\bar \xi}^\alpha(n)\cdot d\xi^\alpha(n)
-d{\bar \xi}^\alpha(n)\cdot\xi^\alpha(n)+{\cal O}((d\xi^\alpha(n))^2)}
{1+\mid\xi^\alpha(n)\mid^2}\right)\\
&\approx &\exp\left\{\sum_\alpha\sum_{n=1}^{M+1}\eps\left[
\frac{J_\alpha}{1+\mid\xi^\alpha(n)\mid^2}
\left(\frac{d{\bar \xi}^\alpha(n)}{\eps}\cdot\xi^\alpha(n)-
{\bar \xi}^\alpha(n)\cdot\frac{d\xi^\alpha(n)}
{\eps}\right)\right]\right\}\nn
\end{eqnarray}
Hence in the limit $M\ra\infty$,
\beq
Z=\int D\mu({\bar \xi},\xi)\exp\left\{i\sum_\alpha
\int_{t_i}^{t_f}dt\left[i
J_\alpha\frac{({\bar \xi}^\alpha\dot \xi^\alpha-
\dot{\bar \xi^\alpha}\xi^\alpha)}{1+\mid\xi^\alpha\mid^2}
- {\cal H}(Q_\alpha^a)\right]\right\}
\eeq
where ${\cal H}(Q_\alpha^a)=<P,\xi\mid {\hat H}({\hat Q}_\alpha^a)\mid P,
\xi>$. Using the classical limit of operators in
coherent state \cite{sch},
we can easily show that
$<P,\xi\mid {\hat Q}_\alpha^a\mid P,\xi>= Q_\alpha^a({\bar \xi},\xi)$ of
Eq.(\ref{def2}). Then it follows
that ${\cal H}$ is simply equal to classical Hamiltonian $H$ of
Eq.(\ref{path}).
Thus, we proved the equivalence between the coherent state
path integral representation Eq.(\ref{path}) and the operator
representation Eq.(\ref{prop}).

In conclusion, the dynamics of the NACS particles are governed by the
Hamiltonian ${\hat H}$ of Eq.(\ref{prop})
\[ \hat {H} = -\sum_\alpha {1\over m_\alpha}\left(\nabla_{\bar
z_\alpha}\nabla_{z_\alpha}  +\nabla_{z_\alpha}\nabla_{\bar
z_\alpha}\right) \]
\beq
\nabla_{z_\alpha} ={\partial\over \partial z_\alpha}  +{1\over 2\pi
\kappa}\left( \sum_{\beta\not=\alpha}
\hat Q^a_\alpha \hat Q^a_\beta {1\over
z_\alpha -z_\beta}+\hat Q^2_\alpha a_z (z_\alpha)\right)\label{ham}
\eeq
\[\nabla_{\bar z_\alpha} ={\partial\over \partial \bar z_\alpha}\]
where $a_z (z_\alpha)=\lim_{z\rightarrow z_\alpha} 1/(z-z_\alpha)$
and the isospin operators $\hat Q^a$'s satisfy the $SU(N+1)$
algebra, $[\hat Q^a_\alpha,\hat Q^b_\beta] =if^{abc} \hat
Q^c_\alpha \delta_{\alpha\beta}$ upon quantizing the classical algebra
Eq.(\ref{op}). This Hamiltonian $\hat H$ has been suggested also in
ref.\cite{ver}.
The second term and the third term
in $\nabla_{z_\alpha}$ are reponsible for the non-Abelian statistics
and the anomalous spins of the NACS particles respectively \cite{lo}.
The connection
$\nabla_{z_\alpha}$ without the third term is the Knizhnik-Zamolodchikov
(KZ) connection \cite{kz} which has been
extensively studied in association
with  the braid groups \cite{braid} and
the rational conformal field theories \cite{ms}.
For the detailed discussion on the physical properties of the NACS
particles, we suggest to consult refs.\cite{lo,lo2}.
The non-Abelian Chern-Simons quantum mechanics described by
the Hamiltonian
Eq.(\ref{ham}) will be useful in studying the fractional quantum Hall
effect  \cite{moo},
the non-Abelian Aharanov-Bohm effect \cite{ver,lo2,wilwu}, etc.

We conclude this letter with a couple of remarks.
Firstly, in performing the path integral, we
did not consider the quantum
fluctuation of the gauge fields around the classical solution
Eq.(\ref{sol}). If we take into account it either by the background
method \cite{witt} or by the perturbative
method \cite{pisa}, we will find
a shift in the coefficient $\kappa$ of the KZ connection
Eq.(\ref{ham}) by
$\kappa\ra \kappa+\frac{c_v}{4\pi}=\kappa+\frac{N}{4\pi}$.
Secondly, we
may treat the constraint Eq.(\ref{sphere})
which reduces $S^{2N+1}$ to
$CP(N)$ on an equal footing with the Gauss' constraint Eq.(\ref{cons}),
and consider the unreduced Lagrangean
\[L'=L'_{ptl}+L_{int}+L_{C-S}\]
where
\beq
L'_{ptl} = \sum_\alpha\left(-{1 \over 2} m_\alpha \dot{\bf
q}_\alpha^2 +
i J_\alpha({\bar Z}^\alpha\dot Z^\alpha-\dot {\bar Z^\alpha}
Z^\alpha)-\la^\alpha\phi^\alpha\right)
\label{lagt}
\eeq
The Lagrangean $L_{ptl}'$ has an advantage that
the $SU(N+1)$ gauge symmetry is
linearly realized,
\[ Z\ra gZ,\ \ \ \ \  Q\ra gQg^{-1}\]
\beq
A\ra gAg^{-1} +igdg^{-1},\ \ \ \ \ \ \ g\in SU(N+1).\label{gau}
\eeq
It also has $U(1)^{N_p}$ gauge symmetry given by
$\de_{\eps_\alpha} f=\eps_\alpha\{\phi^\alpha,f\}$.
However, the BRST quantization based on the Lagrangean Eq.(\ref{lagt})
requires to introduce extra ghosts corresponding to $U(1)^{N_p}$
symmetry and it makes the quantization procedure cumbersome.

\vskip 0.4in
We would like to thank Professors H. S. Song, Y. M. Cho and C. Lee
for their hospitality during our visit at C.T.P. at S.N.U. T. Lee was
supported in part by the KOSEF and
P. Oh was supported by the KOSEF through
C.T.P. at S.N.U.

\end{document}